\documentclass[trackchanges,twocolumn]{aastex7}
\usepackage{lipsum}
\usepackage{amsmath}
\usepackage{enumitem}

\begin{document}

\title{The Dependency of Bar Formation Timescale on Disk Mass Fraction, Toomre $Q$, and Scale Height}

\author[0000-0001-8962-663X]{Bin-Hui Chen}
\affiliation{Tsung-Dao Lee Institute, Shanghai Jiao Tong University, Shanghai 200240, People’s Republic of China}
\affiliation{Department of Astronomy, School of Physics and Astronomy, Shanghai Jiao Tong University, 800 Dongchuan Road, Shanghai 200240, People’s Republic of China}
\affiliation{State Key Laboratory of Dark Matter Physics, School of Physics and Astronomy, Shanghai Jiao Tong University, Shanghai 200240, People's Republic of China}
\email[show]{2000cbh@sjtu.edu.cn}

\author[0000-0001-5604-1643]{Juntai Shen}
\affiliation{Department of Astronomy, School of Physics and Astronomy, Shanghai Jiao Tong University, 800 Dongchuan Road, Shanghai 200240, People’s Republic of China}
\affiliation{State Key Laboratory of Dark Matter Physics, School of Physics and Astronomy, Shanghai Jiao Tong University, Shanghai 200240, People's Republic of China}
\email[show]{jtshen@sjtu.edu.cn}
\correspondingauthor{Juntai Shen}

\defcitealias{bland_etal_2023}{BH23}
\defcitealias{fujii_etal_2018}{F18}

\def\fdisk{f_\mathrm{disk}}
\def\tbar{\tau_\mathrm{bar}}
\def\thickness{h_z/R_\mathrm{d}}

\def\toleranceOfTauBarErr{13.7\ \mathrm{Gyr}}
\def\tauBarMax{100\ \mathrm{Gyr}}
\def\Sbar0Max{0.1}
\def\remainNum{531}

\def\Afit{1216}
\def\Aerr{158}
\def\Bfit{-0.110}
\def\Berr{0.004}

\def\paras{(\fdisk,\ Q,\ \thickness)}
\def\fastTau{0.20}
\def\fastTauUerr{0.09}
\def\fastTauLerr{0.06}
\def\fastTauShow{\fastTau_{-\fastTauLerr}^{+\fastTauUerr}\ \mathrm{Gyr}}
\def\fastParas{(0.67,\ 0.8,\ 1/11)} 

\def\mildTau{4.04}
\def\mildTauUerr{1.24}
\def\mildTauLerr{1.00}
\def\mildTauShow{\mildTau_{-\mildTauLerr}^{+\mildTauUerr}\ \mathrm{Gyr}}
\def\mildParas{(0.42,\ 1.2,\ 1/8)}

\def\slowTau{12.20}
\def\slowTauUerr{3.37}
\def\slowTauLerr{2.80}
\def\slowTauShow{\slowTau_{-\slowTauLerr}^{+\slowTauUerr}\ \mathrm{Gyr}}
\def\slowParas{(0.35,\ 1.2,\ 1/5)}

\def\extremeTau{149.56}
\def\extremeTauUerr{27.26}
\def\extremeTauLerr{25.20}
\def\extremeTauShow{\extremeTau_{-\extremeTauLerr}^{+\extremeTauUerr}\ \mathrm{Gyr}}
\def\extremeParas{(0.13,\ 2.0,\ 1/5)}

\begin{abstract}

Bars are one of the most prominent galactic structures. The classical swing-amplification theory can qualitatively describe the spontaneous bar instability of stellar disks. Still, it cannot quantify the bar formation process or explain why some disk galaxies do not have a bar. Recent studies found that the bar formation timescale depends exponentially on the disk mass fraction of the host galaxy (dubbed as ``Fujii relation"), but they only explored a limited parameter space, where the physical effects of Toomre $Q$ (local disk stability parameter) and disk scale height of the host galaxies are not fully explored. In this work, we check the robustness of the Fujii relation in a higher-dimensional parameter space of disk mass fraction, Toomre $Q$, and scale height. We find that the Fujii relation holds for disk galaxies with physically reasonable Toomre $Q$ and scale height. Furthermore, the bar formation timescale also approximately linearly depends on both Toomre $Q$ and scale height, with a more prolonged bar formation in a hotter or thicker disk. We propose an empirical relation to combine the dependency of the bar formation timescale on the three parameters. Based on the empirical relation and recent observations, we estimate that the bar formation timescale in pure stellar disks ranges from $\fastTauShow$ to $\slowTauShow$ or even significantly beyond the Hubble timescale in some extreme cases.

\end{abstract}


\section{Introduction} \label{sec: intro}

As one of the most prominent galactic structures, bar exists in about 2/3 of disk galaxies \citep[e.g.,][]{eskrid_etal_2000, menend_etal_2007, barazz_etal_2008, sheth_etal_2012, simmon_etal_2014, erwin_2018, lee_etal_2019}. These barred disk galaxies constitute a main branch of the Hubble Sequence \citep{hubble_1926, devauc_1959}. Thus, thoroughly understanding the formation and evolution mechanisms of bar structure is a crucial step in unraveling the origin of the Hubble Sequence, which is one of the most important problems in modern astronomy. As one of the main contributors to the host galaxy's central potential, bar structure has an essential impact on its host galaxy \citep[e.g.,][]{hohl_1971, lyn_kal_1972, combes_etal_1990, athana_1992, kor_ken_2004, athana_2005, min_fam_2010, portai_etal_2015, li_etal_2016, li_etal_2017, fragko_etal_2019, gadott_etal_2020, sai_cat_2022, neuman_etal_2024, chen_etal_2024}, therefore, thoroughly understanding the formation and evolution mechanisms of bars is also invaluable in studying galactic physics.

The prevailing theoretical framework for bar formation is the swing-amplification theory (see \citealt{binney_2020} for a comprehensive introduction). By solving the linearized equations for a razor-thin stellar disk, \cite{toomre_1981} extended the Lin–Shu density wave theory \citep{lin_shu_1964} beyond the tightly wound approximation. In this picture, bars form via a bar instability: transient spiral arms, seeded by noise spectrum, are amplified into trailing spirals through swing-amplification driven by the disk’s differential rotation. These arms are subsequently inverted—likely via reflection at the Lindblad resonances or refraction near the galactic center \citep{bin_tre_2008, sel_mas_2022}—to enter the next swing-amplification, thus establishing a feedback loop. The resulting amplification resembles the resonance in a radio feedback circuit, ultimately leading to the formation of a central bar as the amplified spirals collapse under self-gravity. Alternative mechanisms, such as the groove mode \citep{sel_lin_1989, sel_kah_1991}, have also been proposed. However, the origin of the initial spiral arms is generally attributed to the local dynamical instabilities of the stellar disk, often characterized by Toomre’s $Q$ parameter \citep{toomre_1964}:
\begin{equation}
Q(R)=\dfrac{\sigma_R(R)\kappa(R)}{3.36G\Sigma(R)}.
\end{equation}
Bar formation is possible in disks with $1.0 \lesssim Q \lesssim 2.0$. During the swing-amplification phase, the bar strength,
\begin{equation}
A_2\equiv\left\vert\dfrac{\sum_j m_j\exp(2i\phi_j)}{\sum_j m_j}\right\vert,
\end{equation}
exhibits exponential growth with time, similar to the evolution of an unstable mode \citep{binney_2020, bland_etal_2023}.

However, the swing-amplification theory can only qualitatively describe the bar formation process in an ideal situation. It cannot quantify the bar formation process in real disk galaxies or explain why there are some unbarred disk galaxies \citep{zakhar_etal_2023, sel_car_2023} such as the M33 \citep{sellwo_etal_2019}. Therefore, quantitative studies of bar formation mechanisms beyond the scope of swing-amplification theory frequently employ numerical simulations, which have shed more light on the bar formation mechanism. For example, many \textit{N}-body simulations have verified a bar can form through the bar instability even in a disk with finite thickness \citep[e.g.,][]{com_san_1981, efstat_etal_1982, combes_etal_1990, raha_etal_1991, shen_etal_2010}; \cite{athana_2002} found that an initially dominant dark matter (DM) halo strongly delays the onset of the bar instability.

Recently, \citet[hereafter \citetalias{fujii_etal_2018}]{fujii_etal_2018} found that in \textit{N}-body simulations, the bar formation is somewhat inevitable for disk galaxies in an isolated environment, where the bar formation timescale and age of a disk galaxy determine the galaxy's superficial barred or unbarred appearance. By defining the bar formation timescale $t_\mathrm{bar}$ as the first time when $A_\mathrm{2, max}\equiv\max\{A_2\text{ among radial bins}\}$ exceeds 0.2, they found in their models that $t_\mathrm{bar}$ exponentially depends on the disk mass fraction
\begin{equation}
\fdisk \equiv \dfrac{V_{\mathrm{c,disk}}^2}{V_{\mathrm{c,tot}}^2}\bigg|_{R=2.2R_{\mathrm{d}}}
\end{equation} ($R_\mathrm{d}$ is the scale length of the disk), in the form
\begin{equation}\label{eq: Fujii formula}
t_{\mathrm{bar}}/\mathrm{Gyr}=(0.146 \pm 0.079)\exp \left[(1.38 \pm 0.17) / \fdisk\right].
\end{equation} Later, \citet[hereafter \citetalias{bland_etal_2023}]{bland_etal_2023} proposed a more physical definition of the bar formation timescale $\tbar$ as
\begin{equation}\label{eq: definition of tau bar}
A_2(t) = A_2(0)\cdot\displaystyle{\exp(\frac{t}{\tbar})},
\end{equation} employing $A_2$'s exponential growth nature \citep{binney_2020}. \citetalias{bland_etal_2023} verified the exponential relation between $t_\mathrm{bar}$ and the $\fdisk$. They named it the ``Fujii relation". \citetalias{bland_etal_2023} further proposed an alternative form of the Fujii relation that can match their ``high mass" models:
\begin{equation}\label{eq: BH formula}
\tbar= \tau_{\mathrm{H}}\cdot\exp \left[-13.0\left(f_{\text {disk }}- 0.328\right)\right],
\end{equation}
where $\tau_\mathrm{H}$ is the Hubble timescale. Besides the Fujii relation, they found that the bar formation timescale also depends on the DM halo mass or gas fraction: A greater halo mass, associated with a higher halo concentration in their models, or a greater gas fraction can delay the bar formation. The physical effect of gas fraction on bar formation is further explored later in \cite{bland_etal_2024, bland_etal_2025}.

\citetalias{fujii_etal_2018} and \citetalias{bland_etal_2023} revealed that the bar formation timescale in disk galaxies depends on the disk mass fraction. However, their studies only explored a limited parameter space with few values of Toomre $Q$ and scale height $h_z$. $Q$ and $h_z$ are also important for bar formation: Theoretically, $h_z$ will affect the applicability of the razor-thin disk assumption in the swing-amplification theory; Toomre $Q$ determines the formation rates of transient spiral arms, which are necessary for the bar instability. Besides, studies both in simulations and observations showed that a hotter disk (with a greater Toomre $Q$) will delay the bar formation \citep{ath_sel_1986, sheth_etal_2012, worrak_2025}. Thus, whether the Fujii relation holds in greater parameter space and how $Q$ and $h_z$ affect bar formation timescale still require further exploration.

In this study, we construct several \textit{N}-body models of disk galaxies with different physically reasonable $\fdisk$, $Q$, and $h_z$ (in this paper, we always refer to $Q|_{R=2.2R_{\mathrm{d}}}$ and the exponential scale height). We check the robustness of the Fujii relation in these models and further explore the dependency of the bar formation timescale on $Q$ and $h_z$. We find that the Fujii relation holds in the explored parameter space at fixed $Q$ and $h_z$, and the bar formation timescale also linearly depends on $Q$ and $h_z$. Combining these dependencies, we propose an empirical relation between $\tbar$ and $(\fdisk,\ Q,\ h_z)$. With this empirical relation and recent observations, we estimate the range of the bar formation timescale in pure stellar disks.

The structure of this paper is as follows: In Section \ref{sec: model}, we explain how we set up the \textit{N}-body models. In Section \ref{sec: results}, we present the main results and discuss their implications. Finally, we summarize this paper in Section \ref{sec: summary}

\begin{table}[htbp!]
\caption{Main parameters of the \textit{N}-body models}\label{table: parameters}
\centering
\begin{tabular}{ccccc}
\hline
Component & $N_\mathrm{p}$ & $M/10^{10} M_\odot$ & $a/\mathrm{kpc}$ & $r_\mathrm{c}/\mathrm{kpc}$
\\
DM halo & $500,000$ & 50.0 & 20.0 & 250.0
\\
\hline
Component & $N_\mathrm{p}$ & $R_\mathrm{d}/\mathrm{kpc}$ & $R_{\sigma_R}/\mathrm{kpc}$ 
\\
Stellar disk & $500,000$ & 3.0 & 6.0
\\
\hline
\end{tabular}
\end{table}

\section{Model Setup}\label{sec: model}

We set up the models using the \texttt{AGAMA} package \citep{vasili_2019}. Each model contains a DM halo and a stellar disk. The DM halo has an Navarro–Frenk–White (NFW) profile \citep{navarr_etal_1996}
\begin{equation}\label{eq: halo}
    \rho_\mathrm{NFW}(r)=\rho_0(\frac ra)^{-1}(1+\frac ra)^{-2}\times\exp{[-(\frac r{r_\mathrm{c}})^2]},
\end{equation} where $\rho_0$ is a normalization constant determined by the total mass of the DM halo. The stellar disk has a quasi-isothermal distribution function \citep{vasili_2019}
\begin{equation}\label{eq: disk DF}
\begin{aligned}
f(\boldsymbol{J})&=\dfrac{\tilde{\Sigma}\Omega}{2\pi^2 \kappa^2}\times\dfrac{\kappa}{\tilde{\sigma}_R^2}\exp\left(-\dfrac{\kappa J_R}{\tilde{\sigma}_R^2}\right) \\
&\times \dfrac{\nu}{\tilde{\sigma}_z^2}\exp\left(-\dfrac{\nu J_z}{\tilde{\sigma}_z^2}\right)\times\left.\left\{\begin{array}{ll}1&\text{if }J_\phi\geq0,\\\exp\left(\dfrac{2\Omega J_\phi}{\tilde{\sigma}_R^2}\right)&\text{if }J_\phi<0,\end{array}\right.\right.
\end{aligned}
\end{equation} where $\kappa$/$\nu/\Omega$ is the epicycle/vertical/circular frequency, $J_R,\ J_\varphi,\ \ J_z$ are the actions, and 
\begin{equation}
\begin{split}
\tilde{\Sigma}(R) &= \Sigma_0 \exp{(-R/R_\mathrm{d})}, \\
\tilde{\sigma}_R^2(R) &= \sigma_{R,0}^2\exp{(-2R/R_{\sigma_R})}, \\
\tilde{\sigma}_z^2 (R) &= 2 h_z^2 \nu^2(R).
\end{split}
\end{equation}
More details of the model setup are available in \cite{vasili_2019} and \cite{tepper_etal_2021}.

Because \texttt{AGAMA} uses an iteration strategy to make the model get into equilibrium, the parameters of the created \textit{N}-body model may slightly deviate from the specified parameters \citep{vasili_2019}. Thus, to get a model with the required parameters, we check the actual parameters of each created \textit{N}-body model and modify its specified parameters until the model's actual parameters reach the required values. The steps are as follows:
\begin{enumerate}[itemsep=0pt, topsep=0pt, parsep=0pt]
\item Construct a fiducial model and measure its $\fdisk$, $Q$ and $h_z$.
\item Based on the required values of $(\fdisk,\ Q,\ h_z)$ and the gotten values, we fine-tune the following three specified parameters: first modify the disk mass to match the required $\fdisk$, then modify $\sigma_{R,0}$ to match the required $Q$ and finally modify the disk scale height to match the required $h_z$.
\item Construct the fine-tuned model, and measure its $\fdisk$, $Q$ and $h_z$.
\item Repeat the above two steps until the $\fdisk$, $Q$, and $h_z$ of the created model converge to the desired values within some tolerances.
\end{enumerate}

Using a semi-automated program implementing the above steps, we construct 693 \textit{N}-body models with $\fdisk=$ 0.30, 0.35, ..., 0.80 (11 values); $Q=$ 0.8, 1.0, ..., 2.0 (7 values); $h_z/\mathrm{kpc}=$ 0.2, 0.3, ..., 1.0 (9 values). The convergent tolerances are $\epsilon(\fdisk)=0.025$, $\epsilon(Q)=0.1$, and $\epsilon(h_z)=0.05\ \mathrm{kpc}$. Table \ref{table: parameters} lists other parameters of these models. The disk scale length is fixed, so $h_z/R_\mathrm{d}$ ranges from $6.7\%$ to $33.3\%$ in the models. To facilitate the exploration of the great parameter space, we adopt a slightly low particle resolution in both the disk and halo, which may introduce collisional relaxation/heating \citep{sellwo_2013, ludlow_etal_2021, wilkin_etal_2023}. In the Appendix~\ref{app:indep}, we verify that our main results on bar formation timescales are robust when the particle resolution is doubled.

Note that the initial condition of the models includes no bulges. Though bulges are also one of the most prominent structures in disk galaxies, most bulges emerge from the formed bar \citep{she_zhe_2020}, so they might not exist in the precursor disks. Other spheroidal classical bulges are dynamically similar to a compact DM halo, which reduces $\fdisk$ by increasing the central mass concentration of the non-disk ingredient. However, as the mass of classical bulge in Milky Way–like disk galaxies is $\lesssim10\%$ of the disk \citep{shen_etal_2010}, its impact on bar formation is not significant, so we neglect it to simplify the models.

We use \texttt{GADGET4} code \citep{spring_etal_2021} to evolve the created \textit{N}-body models. Each model is evolved for an adaptive time ranging from $8\ \mathrm{Gyr}$ to $40\ \mathrm{Gyr}$ so that the disk has enough time to form a bar.

\begin{figure*}[htbp!]
\centering
\includegraphics[width=1.\textwidth]{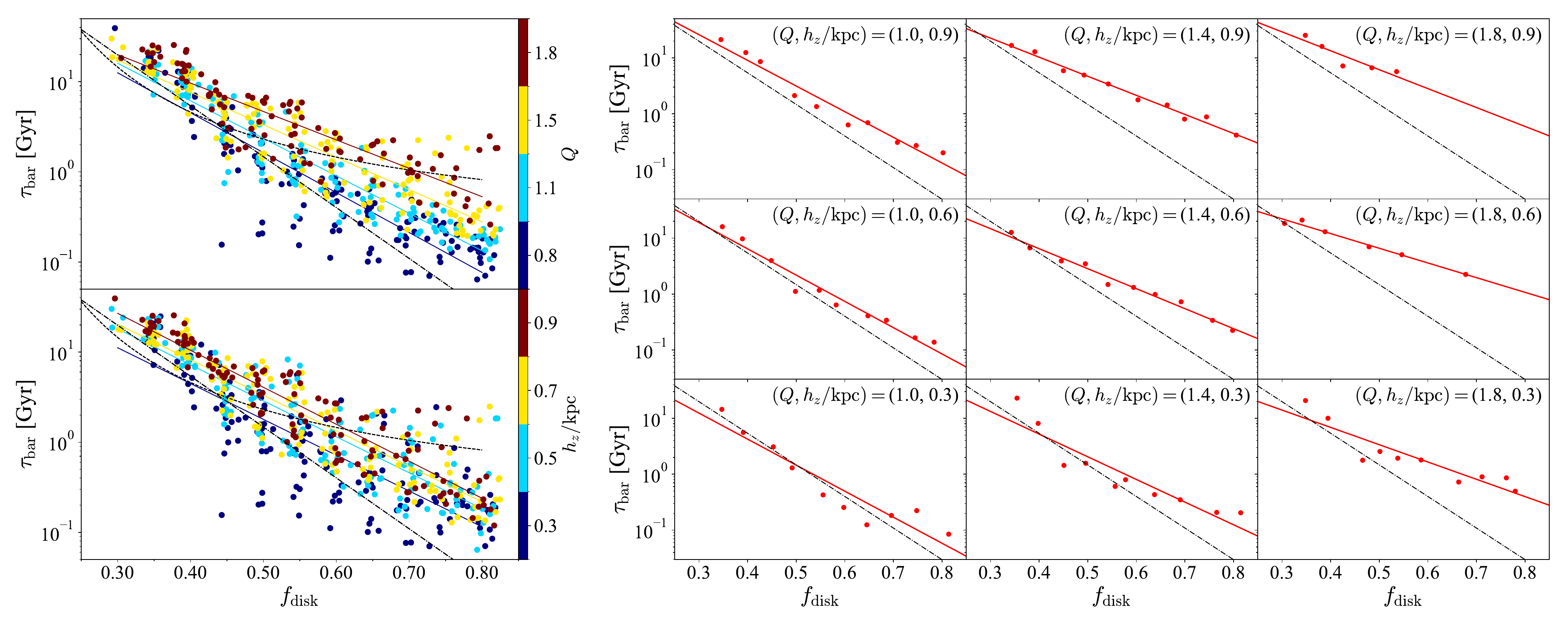}
\caption{Left column: the ``Fujii diagram", namely $\tbar$ against $\fdisk$, for the sample with controlled $Q$ (upper panel) or $h_z$ (lower panel) indicated by the color of the data points. The solid lines are the linear fits of the data points in the corresponding color. For comparison, we also show the Fujii relations reported in \citetalias{fujii_etal_2018} (black dashed curve, Equation \ref{eq: Fujii formula}) and \citetalias{bland_etal_2023} (black dotted-dashed curve, Equation \ref{eq: BH formula}). Right columns: similar Fujii diagrams but for models with several fixed $(Q,\ h_z/\rm{kpc})$; the unshown models just show similar results. Consistent with \citetalias{bland_etal_2023}, $\ln\tbar$ in our sample also linearly anticorrelates with $\fdisk$, shown by the data points' diagonal distribution from top left to bottom right in each panel. Besides the Fujii relation, the $\tbar$ secondarily depends on $Q$ and $h_z$: a hotter/thicker disk corresponds to a slower bar formation.}
\label{fig: Fujii relation}
\end{figure*}

\begin{figure*}[htbp!]
\centering
\includegraphics[width=1.\textwidth]{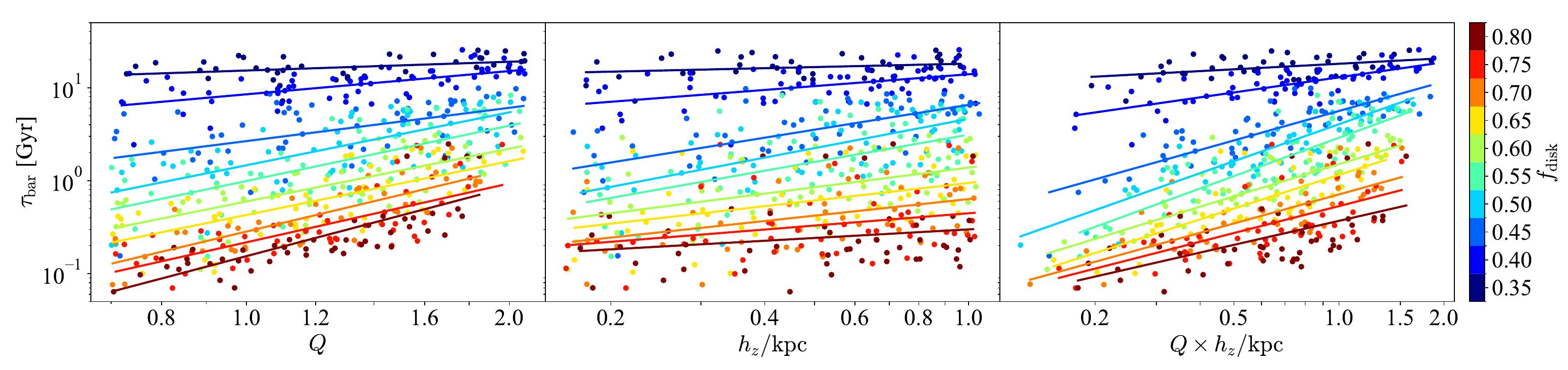}
\caption{The distribution of $\tbar$ against $Q$ (left panel), $h_z$ (middle panel), and $Q\times h_z$ (right panel). The colors represent the models' $\fdisk$, and the solid lines are the linear fits of the data points in the corresponding color. Approximately, $\tbar$ linearly depends on $Q$ and $h_z$ in a similar fashion, and a more prominent linear correlation relative to $Q\times h_z$ is visible in the right panel.}
\label{fig: linear relation}
\end{figure*}

\begin{figure*}[htbp!]
\centering
\includegraphics[width=.9\textwidth]{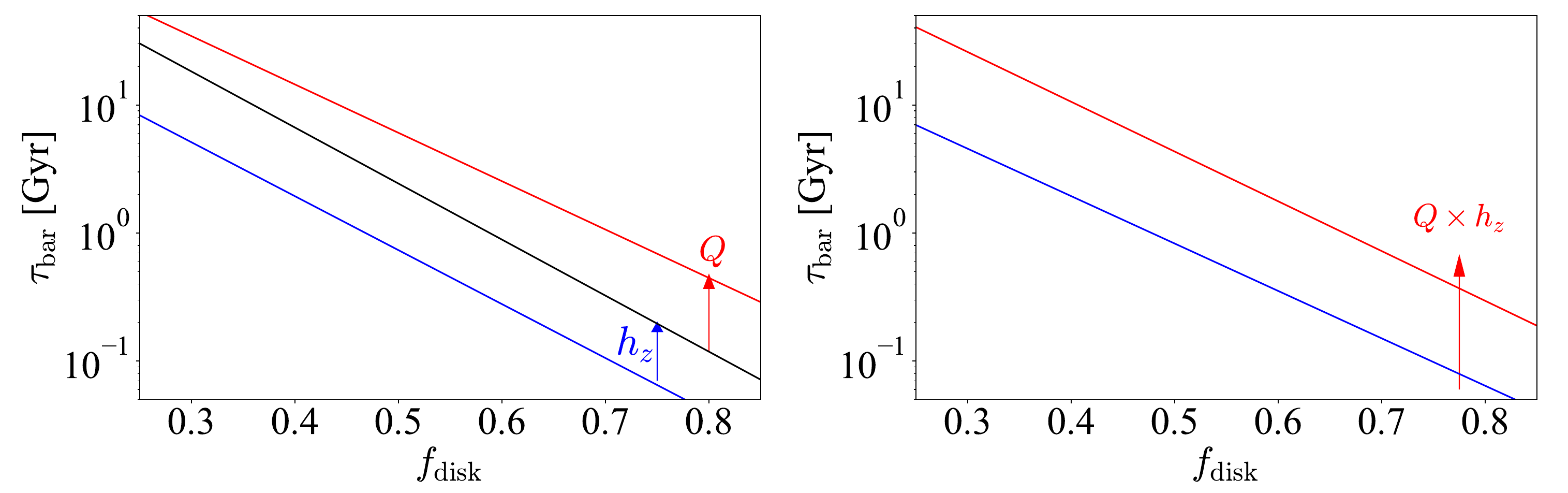}
\caption{To better visualize how the trend between $\tbar$ and $\fdisk$ varies with $Q$ and $h_z$, we further show some results extracted from Figures \ref{fig: Fujii relation} and \ref{fig: linear relation}. Left panel: results with $(Q, h_z/\mathrm{kpc})=$ $(1.0, 0.2)$ in blue, $(1.0, 0.8)$ in black, and $(1.8, 0.8)$ in red. Right panel: similar but for $\tbar$ against $\fdisk$ with $Q\times h_z/\mathrm{kpc}=$ 0.2 in blue and 1.0 in red.}
\label{fig: illustrative}
\end{figure*}

\section{Results and Discussions}\label{sec: results}
\subsection{Distribution of $\tbar$}
Using \texttt{scipy.optimize.curve\_fit} with the default Levenberg-Marquardt algorithm to solve the non-linear least-square problem, we fit the exponential growth part of $A_2(t)$ with Equation \ref{eq: definition of tau bar} to get the bar formation timescale $\tbar$ of each model, similar to \citetalias{bland_etal_2023}. Models with pathological fitting results that $A_2(0)\geq\Sbar0Max$ or $\tbar\geq\tauBarMax$, or fitting error $\Delta(\tbar)\geq\toleranceOfTauBarErr$ are excluded. This typically excludes models with nearly zero $A_2$ throughout the run, as their bar formation timescales are too long to be resolved within the duration of the simulation. There remain $\remainNum$ models, which we refer to as the sample hereafter.

Figure \ref{fig: Fujii relation} displays the distribution of $\tbar$ against $\fdisk$ in the sample, for both the whole sample in the left column and models with fixed $(Q,\ h_z)$ in the right columns. In each panel, the solid curves are the linear fits of the data points in the corresponding color. They all distribute around the top left to bottom right diagonal, indicating a linearly negative correlation between $\ln\tbar$ and $\fdisk$. Besides, the range of $\tbar$ is consistent with \citetalias{fujii_etal_2018} and \citetalias{bland_etal_2023}. Thus, the Fujii relation also holds in disk galaxies with $0.8\lesssim Q\lesssim 2.0$ and $0.2\lesssim h_z/\mathrm{kpc}\lesssim 1.0$. Notably, in the left column of Figure \ref{fig: Fujii relation}, $\tbar$ also depends on $Q$ and $h_z$, where a greater $Q$ or $h_z$ corresponds to a delayed bar formation.

To delve deeper into the relation between $\tbar$ and $(Q,\ h_z)$, in Figure \ref{fig: linear relation}, we show the distribution of $\tbar$ against $Q$ (left panel) and $h_z$ (middle panel). The colors indicate the models' $\fdisk$ and the solid straight lines are the linear fits of the data points in the corresponding color. The left two panels of Figure \ref{fig: linear relation} show that $\tbar$ approximately linearly increases with $Q$ and $h_z$ for fixed $\fdisk$ (or more precisely, a power law with exponent close to $1$). Notably, the increasing trends of $\tbar$ against $Q$ and $h_z$ are similar, indicating their degeneracy in shaping the $\tbar$. Thus, in the right panel of Figure \ref{fig: linear relation}, we also show the distribution of $\tbar$ against $Q\times h_z$, which shows that there is a similar linear relation between $\tbar$ and $Q\times h_z$.

In summary, the $\tbar$ exponentially depends on $\fdisk$ and linearly depends on $Q\times h_z$. Figure \ref{fig: illustrative} illustrates these two trends.

\begin{figure*}[htbp!]
\centering
\includegraphics[width=1.\textwidth]{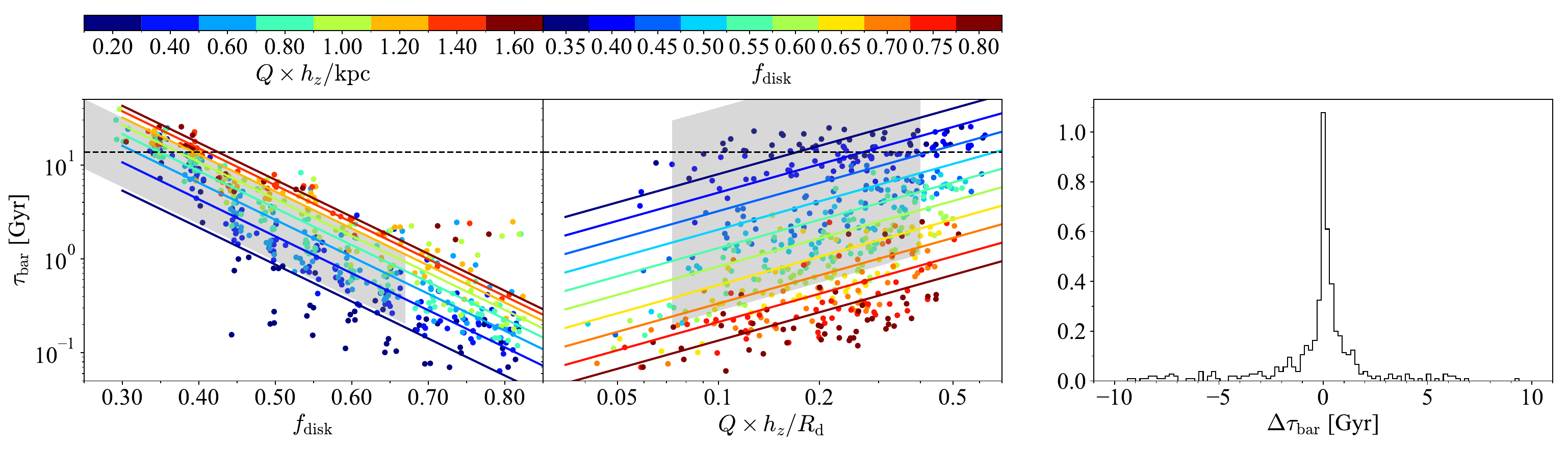}
\caption{Left two panels: the comparison between the measured $\tbar$ (data points) and the predicted values of Equation \ref{eq: fit} (solid lines). Globally, the measured $\tbar$ aligns with Equation \ref{eq: fit}. In both panels, we use black dashed curves to represent the locus of the Hubble timescale and gray shadows to represent the range of the bar formation timescale for pure stellar disks in the Universe (see more details in Section \ref{sec: results: constrain}). Right panel: the distribution of $\Delta\tbar\equiv\tau_\mathrm{bar, predicted} - \tau_\mathrm{bar, measured}$, where $\tau_\mathrm{bar, predicted}$ is calculated with Equation \ref{eq: fit} based on the models' $\fdisk$, $Q$, and $h_z$. Note that in all runs $R_\mathrm{d}=3.0\ \mathrm{kpc}$.}
\label{fig: verification}
\end{figure*}

\subsection{Empirical relation for $\tbar$, $\fdisk$, $Q$ and $h_z$}
The above results show that $\tbar$ depends exponentially on $\fdisk$ and approximately linearly on $Q$ and $h_z$. We find that these trends are unaffected if we replace the exponential bar formation timescale in Equation \ref{eq: definition of tau bar} with $t_\mathrm{bar}$ as the first time when $A_2$ first exceeds some threshold value, as done by \citetalias{fujii_etal_2018}. The $\tbar$ varies slightly if we increase the softening length by a factor of 2, and by a factor of 7 it systematically increases about $20\%$ as expected \citep{sel_ath_1986}. Still, the dependencies of $\tbar$ on $\fdisk$ and $Q\times h_z$ persist. Thus, these dependencies are physical results rather than numerical artifacts. 

These dependencies can be combined as an empirical relation as follows:
\begin{equation}
\label{eq: empirical}
\dfrac{\tbar}{[T]} = A Q\dfrac{h_z}{R_\mathrm{d}}\exp(\dfrac{\fdisk}{B}),
\end{equation} where $[T]=\mathrm{1.0\ Gyr}$ is the simulation's unit of time and $R_\mathrm{d}=3.0\ \mathrm{kpc}$ in all runs. We include $[T]$ and use the dimensionless relative thickness $\thickness$ rather than $h_z$, aiming to make the free parameters $A$ and $B$ dimensionless so that the above formula has a more general applicability. Using \texttt{scipy.optimize.curve\_fit} again, we fit the distribution of $\tbar$ in the sample with the above formula and get
\begin{equation}
\label{eq: fit}
\dfrac{\tbar}{\Afit\ \mathrm{Gyr}} = Q\dfrac{h_z}{R_\mathrm{d}}\exp(-\dfrac{\fdisk}{0.110}),
\end{equation} where the fitting results are $A=\Afit\pm\Aerr$ and $B=\Bfit\pm\Berr$. Neglecting the fitting errors, we can equivalently write the above equation as
$\tau_{\text {bar }} / \mathrm{Gyr}=Q \frac{h_z}{R_{\mathrm{d}}} \exp (-\frac{f_{\text {disk }}}{0.11}+7.1)$.

In Figure \ref{fig: verification}, we compare the measured $\tbar$ and the ones predicted by Equation \ref{eq: fit}. The left two panels show the distribution of $\tbar$ against $\fdisk$ and $Q\times h_z/R_\mathrm{d}$, where the solid lines, calculated from Equation \ref{eq: fit}, are roughly consistent with the measured $\tbar$. The right panel shows the distribution of differences between the predicted and measured $\tbar$. The distribution concentrates sharply around $0$, showing the effectiveness of the above empirical relation. Thus, Equation \ref{eq: fit} can quantitatively predict the variation of $\tbar$ in the three-dimensional parameter space of $(\fdisk,\ Q,\ h_z)$.

\subsection{Implication of the empirical relation}

We provide a somewhat speculative explanation attempting to understand the above empirical relation.

The spontaneously formed bars originate from amplified spirals in initially unperturbed stellar disks, through the swing-amplification feedback loop of shearing sheets \citep{binney_2020} or a succession of groove instabilities \citep{sel_car_2019}. \citetalias{bland_etal_2023} found that the initial growth of the bar strength is in a nearly exponential fashion, similar to an unstable mode \citep{binney_2020}. Our models show similar initial exponential growth. Thus, we consider a bar grows in the form of Equation \ref{eq: definition of tau bar}, assuming that the bar strength grows by $g$ during a characteristic timescale $\Delta t$ \footnote{For example, the period of one swing-amplification loop or the timescale of a single groove mode. In the discussion later, $\Delta t$ can actually be defined as an infinitesimally short interval with no connection to any physical timescale. However, binding it to a physical timescale during bar formation is more natural.} at time $t$, namely
\begin{equation}\label{eq: definition of g}
g = \dfrac{A_2(t+\Delta t) - A_2(t)}{A_2(t)}.
\end{equation} Substituting Equation \ref{eq: definition of tau bar} into it, we get
\begin{equation}\label{eq: g and tau}
g = \exp(\dfrac{\Delta t}{\tbar})  - 1.
\end{equation}
The spiral patterns rotate at a rate comparable to the local radial/epicycle frequency $\kappa$ \citep{toomre_1981} during bar formation. Thus we can estimate $\Delta t\sim 1/\kappa$, which is of the order of $0.01\ \mathrm{Gyr}$ in the bar region where $\kappa\gtrsim 90\ \mathrm{km\,s^{-1}\,kpc^{-1}}$ (extrapolated inward from the values for the Galactic outer disk reported by \citealt{lepine_etal_2008}). This timescale is much shorter than the bar formation timescales we found, so we can approximate the above equation with its linear expansion
\begin{equation}
g = \dfrac{\Delta t}{\tbar}.
\end{equation}
Thus
\begin{equation}\label{eq: intermediate}
\tbar = \dfrac{\Delta t}{g},
\end{equation} where $\Delta t$ and $g$ on the right-hand side may vary with time, but their ratio always equals $\tbar$.

\cite{toomre_1981} found that the logarithm of ``maximum full-swing amplification factor" is tightly correlated with 
\begin{equation}
X =\frac{\kappa^2 R}{2 \pi G \Sigma m}
\end{equation} (Equation 6.77 of \citealt{bin_tre_2008}), where for a Mestel disk ($V_c=\mathrm{constant}$) $X=1/\fdisk$ for an $m=2$ mode. As shown in Figure 7 of \cite{toomre_1981}, for a razor-thin disk model \citep{zang_1976}, the logarithm of the maximal growth factor is a sharply decreasing function of $X$ when $X\gtrsim1.25$ (corresponding to $\fdisk\lesssim 0.8$). Inspired by their results, we speculate that in our models we may approximate $\ln g$ as
\begin{equation}\label{eq: ansatz of g}
\ln g\simeq \dfrac{C}{X} = C\fdisk,
\end{equation} for some positive constant $C$. The above form is no more than an ansatz, but it can help us understand how the bar formation timescale correlates with $\fdisk$.

The classical swing-amplification theory indicates that the characteristic timescale of a spiral pattern should also depend on $Q$ and $h_z$. (1) A disk with a lower $Q$ has a greater wavelength range of unstable density wave modes. Thus, a lower $Q$ corresponds to a lower $\Delta t$. Notably, the classical theory forbids the spiral instability in a disk with $Q>1$, which is inconsistent with the results of \textit{N}-body simulations in the literature. This reflects the necessity of modification on the classical Toomre $Q$ criterion (e.g., \citealt{behren_etal_2015, geo_sch_2025}). However, a thorough discussion of this question is beyond the scope of this work. (2) Finite $h_z$ breaks the razor-thin assumption of the classical theory, which changes the dynamical timescale (Equation 2.40 of \citealt{bin_tre_2008}) as 
\begin{equation}
t_\mathrm{dyn}\simeq\dfrac{1}{\sqrt{G\rho}} \propto \sqrt{\dfrac{\pi R^2 h_z}{M}}.
\end{equation} This reflects that a thicker disk should have a longer $\Delta t$. Therefore, $\Delta t$ should be some increasing function of both $Q$ and $h_z$:
\begin{equation}\label{eq: delta t ansatz}
\Delta t = f\left(Q,\ \dfrac{h_z}{R_\mathrm{d}}\right).
\end{equation}

Substituting Equations \ref{eq: ansatz of g} and \ref{eq: delta t ansatz} into Equation \ref{eq: intermediate}, the bar formation timescale should have a form as 
\begin{equation}\label{eq: final}
\tbar = f\left(Q,\ \dfrac{h_z}{R_\mathrm{d}}\right) \exp(-C\fdisk),
\end{equation} which is consistent with Equation \ref{eq: fit}. Since theorists are still struggling with an analytic theory of bar formation, we cannot rigorously derive the functional form of $f(Q,\ {h_z}/{R_\mathrm{d}})$. Our numerical experiments seem to prefer the linear form of $Q$ and $h_z$,
\begin{equation}\label{eq: final result}
\tbar \propto Q\times \dfrac{h_z}{R_\mathrm{d}} \exp{(-C \fdisk)}.
\end{equation}

In summary, the empirical relation may stem from two aspects: the growth rate depends exponentially on $\fdisk$ when $\fdisk\lesssim0.8$ (consistent with \citealt{bland_etal_2023, bland_etal_2024}), and the disk's hotness and thickness secondarily, but still significantly, increase the timescale of bar growth. In the presence of gas, the analysis for stellar disks may need to be modified \citepalias{bland_etal_2023}, as there are more complex physics \citep{bland_etal_2024, bland_etal_2025}. We also refer interested readers to the detailed discussion in \cite{bland_etal_2025} about the JT66 shearing sheet approach \citep{jul_too_1966, binney_2020}.

\subsection{Constraining $\tbar$ in pure stellar disks}\label{sec: results: constrain}
Equation \ref{eq: fit} may help us to constrain the bar formation timescale of pure stellar disks, with presupposed ranges of $\fdisk$, $Q$, and $\thickness$: (1) \cite{price_etal_2021} presented a systematic analysis of 41 massive, large star-forming galaxies at cosmological redshift $0.67\lesssim z\lesssim 2.45$. Their disk mass fractions range from $0.12$ to $0.67$ and concentrate around $0.45$ (see Figure 6 of \citetalias{bland_etal_2023}). We employ their reported range of $\fdisk$. (2) There are few direct observations on the $Q$ of external galaxies, and we assume $Q$ of disk galaxies ranges from $0.8$ to $2.0$, covering values from unstable to stable regimes. (3) Some recent observations show that the scale height of disk galaxies in the Universe ranges approximately from $0.1\ \mathrm{kpc}$ to $1\ \mathrm{kpc}$ \citep{lia_luo_2024, ranaiv_etal_2024, tsukui_etal_2025}, with the relative thickness $\thickness$ ranging from $1/5$ to $1/11$ which is independent of redshift when $0.1<z<3.0$ \citep{tsukui_etal_2025}. With these ranges and Equation \ref{eq: fit}, we estimate the range of the bar formation timescale in pure stellar disks, shown as gray shadows in Figure \ref{fig: verification}. Among this range, some characteristic values are as follows: the fastest bar formation in the Universe has $\tbar=\fastTauShow$ for $\paras = \fastParas$, a milder bar formation has $\tbar=\mildTauShow$ for $\paras = \mildParas$, and a slow bar formation has a timescale $\tbar=\slowTauShow$ closing to a Hubble timescale for $\paras = \slowParas$, consistent with \citetalias{fujii_etal_2018} and \citetalias{bland_etal_2023}.

Note that in real disk galaxies, the presence of gas can further delay bar formation, with the degree of delay depending on the gas mass fraction (\citetalias{bland_etal_2023}; \citealt{bland_etal_2024, bland_etal_2025}). Therefore, the preceding estimates should be regarded as the lower limit on the bar formation timescale in real disk galaxies.

\section{Summary} \label{sec: summary}
In this paper, we extend the Fujii relation of bar formation timescale in disk galaxies to a three-dimensional parameter space of $(\fdisk,\ Q,\ h_z)$, where the bar formation timescale also depends approximately linearly on $Q$ and $h_z$. We proposed an empirical relation (Equation \ref{eq: fit}) to characterize these dependencies, and found that such a relation can be motivated by the classical theory. With the recent observations, we estimate that the bar formation timescale for pure stellar disks in the Universe is about $\fastTauShow$ for the coldest and most massive disks, $\mildTauShow$ for disks with milder parameters, and close to or even exceeding a Hubble timescale for extremely light/hot/thick disks.

\begin{acknowledgments}
We thank the referee for the careful review and stimulating suggestions. We sincerely thank Thor Tepper-Garcia and Joss Bland-Hawthorn for the help with the initial condition setup. We also sincerely thank Jerry A. Sellwood, Glenn van der Ven, Ling Zhu, Zhao-Yu Li, Sandeep K. Kataria, and Tigran Khachaturyants for the helpful discussions. Bin-Hui Chen gratefully acknowledges the financial support from the China Scholarship Council and the support of the T.D. Lee scholarship. The research presented here is partially supported by the National Natural Science Foundation of China under grant Nos. 12025302, 11773052, and 11761131016; by China Manned Space Program with grant no. CMS-CSST-2025-A11; and by the “111” Project of the Ministry of Education of China under grant No. B20019. This work used the Gravity Supercomputer at the Department of Astronomy, Shanghai Jiao Tong University, and the facilities in the National Supercomputing Center in Jinan.
\end{acknowledgments}

\appendix
\section{Convergence Test of Particle Resolution} \label{app:indep}

\begin{figure}[htbp!]
\centering
\includegraphics[width=.48\textwidth]{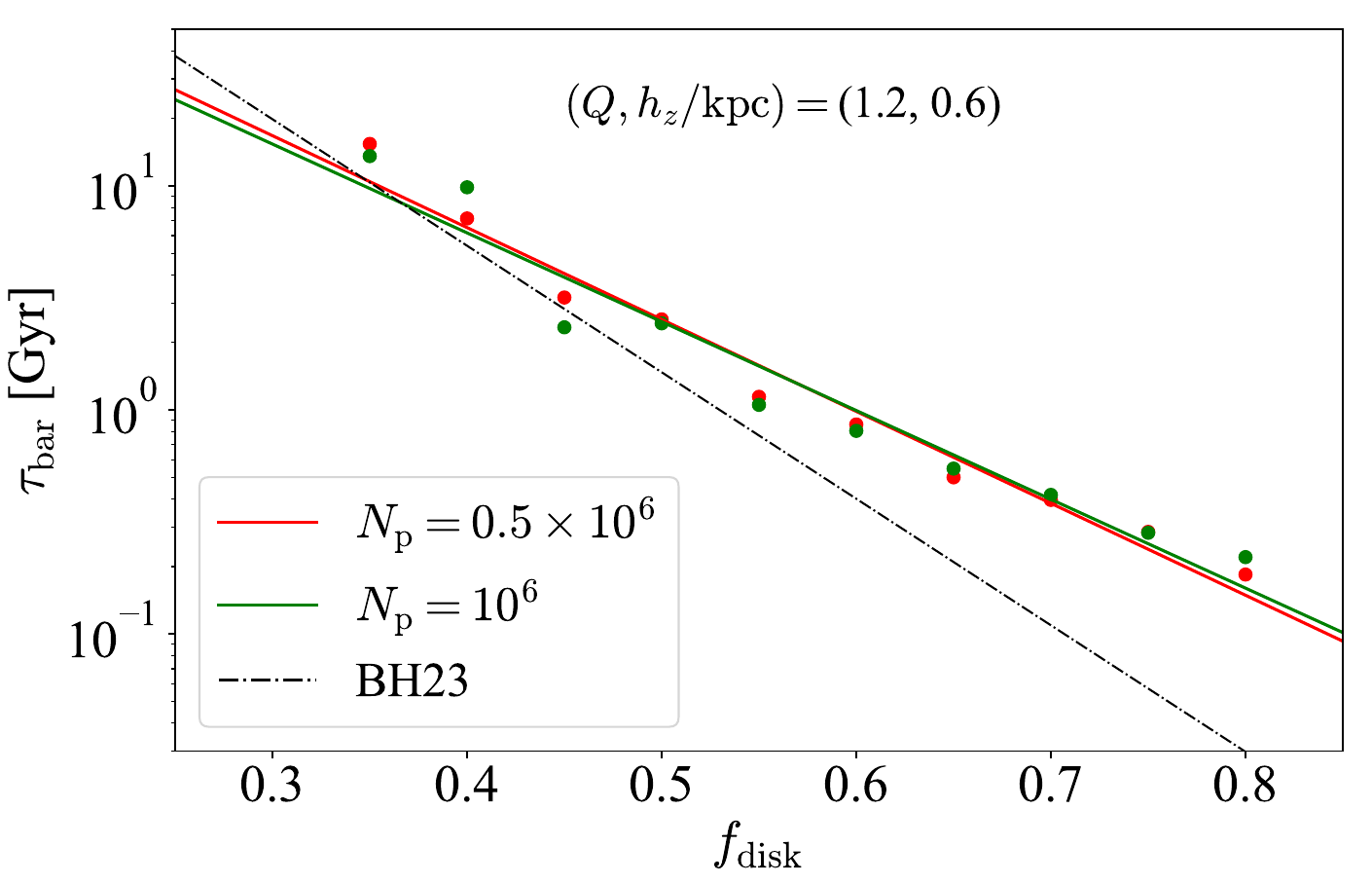}
\caption{Fujii diagram for models with $(Q,\ h/\mathrm{kpc}) = (1.2,\ 0.6)$. Red points are taken from the main text, while green points correspond to models with twice the particle number in both the halo and disk. Colored solid lines show linear fits to the respective data points. For comparison, we also plot the relation reported by \citetalias{bland_etal_2023} (black dotted-dashed line, Equation \ref{eq: BH formula}). The higher-resolution models are consistent with the lower-resolution ones.}
\label{fig: independent test}
\end{figure}

To assess whether the slightly low particle number $N_\mathrm{p}$ used in the main results affects the relation between the bar formation timescale and $(\fdisk,\ Q,\ h_z)$, we constructed models for $(Q,\ h/\mathrm{kpc}) = (1.2,\ 0.6)$ with doubled $N_\mathrm{p}$ in both the disk and halo. The results are shown in Figure~\ref{fig: independent test}. The higher-resolution models are consistent with those used in the main text. This confirms that the derived bar formation timescales are not significantly impacted by the slightly low resolution adopted in the main text, which is exploited to facilitate the exploration of the three-dimensional parameter space.

\bibliography{references}{}
\bibliographystyle{aasjournalv7}

\end{document}